\documentclass[a4paper]{article}

\usepackage{icrc2013}
\usepackage[english]{babel}

\title{Status of the Monoscopic Analysis Chains for H.E.S.S. II}

\shorttitle{Status of the Monoscopic Analysis Chains for H.E.S.S. II}

\authors{
Markus Holler$^{1,2}$,
Arnim Balzer$^{1,2}$,
Yvonne Becherini$^{3}$,
Stefan Klepser$^{2}$,
Thomas Murach$^{4}$,
Mathieu de Naurois$^{5}$,
Robert Parsons$^{6}$,
for the H.E.S.S. Collaboration.
}

\afiliations{
$^1$ University of Potsdam, Germany \\
$^2$ DESY, Zeuthen, Germany \\
$^3$ University of Heidelberg, Germany \\
$^4$ Humboldt University, Berlin, Germany \\
$^5$ Ecole Polytechnique Paris, France \\
$^6$ Max Planck Institute for Nuclear Physics, Heidelberg, Germany.
}

\email{markus.holler@desy.de}

\abstract{H.E.S.S. is a system of Imaging Atmospheric Cherenkov Telescopes (IACTs) measuring cosmic gamma-rays with very high energies in Namibia. Extending the array with a fifth telescope with a mirror area of $600\,\mathrm{m}^2$ leads to a lower energy threshold as well as an increased sensitivity of the system. Moreover, it is now the first IACT array consisting of telescopes with different sizes. Low-energetic gamma-rays detected by the telescopes can either be analyzed monoscopically, allowing for a lower threshold, or stereoscopically, using hybrid events only which leads to a better reconstruction performance. We present the status of the monoscopic analysis of H.E.S.S. II events. In order to cross-check the results, we use two independent analysis chains, based on different reconstruction methods. The first method uses the second moments of the cleaned camera image (Hillas parameters) in order to deduce the properties of the primary particle. The background discrimination of this method can be optimized with multi-variate analysis techniques. The second method is based on the comparison of the camera image with the results of a semi-analytical model of the air shower using a Loglikelihood-Maximization. We present the status of these analysis efforts and their respective performances. One of the chains has been applied on real data of the Crab Nebula. All results shown here have to be considered preliminary.}

\keywords{$\gamma$-ray astronomy, analysis methods, monoscopic reconstruction, Monte Carlo simulations.}

\begin{document}
\maketitle

\section{Introduction}
\label{intro}

H.E.S.S. is a system of five Imaging Atmospheric Cherenkov Telescopes (IACTs) located in Namibia. Four of the IACTs are arranged on a square with an edge length of $120\,$m, have a mirror diameter of around $13\,$m and are operational since 2004 (H.E.S.S. I). Each of them is equipped with a camera containing $960$ photomultipliers (PMTs). In 2012, the array was expanded by a large-size IACT (mirror size $32\,\mathrm{m} \times 24\,\mathrm{m}$) in the center of the existing array (H.E.S.S. II). In addition to the intensity, the camera of the new telescope also records the Time of Maximum (ToM) and Time over Threshold (ToT) of each of the $2048\,$PMTs. In contrast to H.E.S.S. I, H.E.S.S. II will not only read out stereoscopic events where at least two of the IACTs are triggered by the same shower, but also events that only triggered the new telescope.

In this proceeding, we present the current status of the monoscopic analysis chains for H.E.S.S. II. Section~\ref{reconstruction} contains a brief overview of the monoscopic reconstruction methods currently used in the H.E.S.S. software. In Section~\ref{mono_analysis}, we outline the challenges of the monoscopic reconstruction technique, followed by a description of the expected performance in terms of effective area and integral sensitivity in Section~\ref{performance}. The paper ends with some conclusive remarks in Section~\ref{concl}.

\section{Reconstruction Methods}
\label{reconstruction}

 \begin{figure*}[!t]
  \centering
  \includegraphics[width=\textwidth]{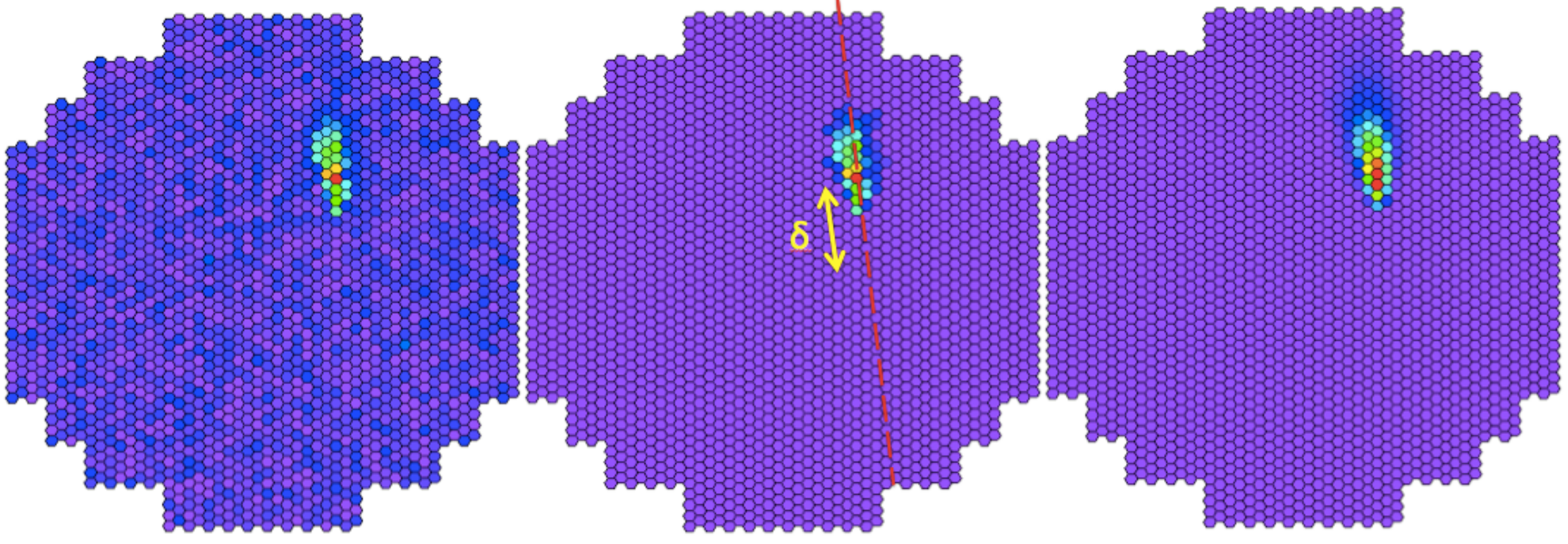}
  \caption{\textit{Left panel:} intensity of a simulated $\gamma$-ray with an energy of $125\,$GeV and an impact distance of $120\,$m in the H.E.S.S. II camera. \textit{Central panel:} same event, but cleaned using the intensity and ToM information. The length axis of the Hillas ellipse is indicated with a dashed red line, whereas the distance from the Center of Gravity of the shower to the direction of the $\gamma$-ray is shown in yellow. \textit{Right panel:} best-fitting model template for the same event.}
  \label{camera_fig}
 \end{figure*}

There are different approaches to deduce the properties of the primary particle from the camera image of an IACT. The H.E.S.S. collaboration uses several independent reconstruction methods, allowing to cross-check results and estimate systematic uncertainties. In the following, we briefly introduce the two methods which are currently used for the monoscopic reconstruction.

\subsection{Hillas Reconstruction}

The standard reconstruction technique in very high-energy (VHE) $\gamma$-ray astronomy is based on the second moments of the cleaned camera image \cite{bib:hillas}, called length and width. Together with the total number of photoelectrons in the cleaned image (called size), it is possible to estimate the arrival direction, energy and impact point on the ground of the particle. The hadronic background is reduced by accepting only events within certain length and width values. For a more detailed description of this reconstruction method as used by H.E.S.S. see \cite{bib:CrabHess}.

Instead of using fixed and manually-set cut values for the $\gamma$/hadron separation, one can make use of machine-learning algorithms like e.g. \textit{Boosted Decision Trees} (BDT) for a better characterization of the event type. Applying such methods on the parameters obtained from a Hillas analysis leads to an improvement of the overall analysis performance, as described in \cite{bib:OhmTMVA}.

\begin{table}[h]
\begin{center}
\begin{tabular}{|l|c|c|}
\hline Analysis Chain & Type & $\gamma$/hadron Separation \\  \hline
1 & Hillas & BDT \\ \hline
2 & Model & Combined Cut \\ \hline
\end{tabular}
\caption{Monoscopic analysis chains used in the H.E.S.S. software. Whereas \textit{Chain 1} is based on machine-learning algorithms, a combined cut is used for \textit{Chain 2} (see text for further details).}
\label{analysis_table}
\end{center}
\end{table}
In the following, the Hillas-based analysis chain is referred to as \textit{Chain 1}. It is advanced compared to a standard Hillas analysis in the way that it uses automated parameter training algorithms as described above.

\subsection{Model Reconstruction}

The second method uses a semi-analytical model of the air shower initiated by a $\gamma$-ray (content of the whole section adapted from \cite{bib:MathieuModel}). Using this model, it is possible to calculate the expected Cherenkov photon density on the ground for a given primary energy, first interaction depth, and impact distance. For a given zenith angle, several thousands of templates are computed, covering sufficiently-sized ranges for each of the parameters. When analyzing an event, the method first performs a standard Hillas analysis in order to get seed parameters. Afterwards the intensity of each pixel is compared to the expected value from the respective model template by calculating a Loglikelihood value. The overall Loglikelihood of all participating telescopes is maximized in order to find the best parameters of the event. The best-fit value is compared to the expected Loglikelihood under the assumption of a $\gamma$-ray (called Goodness) which proved to be a powerful background discrimination variable for H.E.S.S. I.

In order to further improve the background rejection for the case of monoscopic reconstruction, the \textit{Chain 2} presented here uses a combined cut of several parameters. In addition to the Goodness and the ShowerGoodness (Goodness calculated only in the shower region), the Hillas parameters Length and Width are used in order to classify events as $\gamma$-like.

\section{Monoscopic Analysis}
\label{mono_analysis}

Together with the large mirror area of the new telescope of H.E.S.S., the analysis of monoscopic events enables it to detect and reconstruct $\gamma$-rays with much lower energies that do not trigger the smaller telescopes. As was shown by \cite{bib:MagicTiming}, it is possible to use the timing information of the camera for the image cleaning in order to reconstruct faint showers. An example of a camera image which was cleaned using the intensity as well as the ToM information is given in Fig.~\ref{camera_fig} (central panel). Cleaning with the use of the ToM information is currently used for \textit{Chain 2}.

Whereas the direction of a $\gamma$-ray can be well determined when using stereoscopic reconstruction by intersecting the length axes of the Hillas ellipses, it becomes more complicated in the case of monoscopic reconstruction. In order to dissolve the ambiguity of the shower direction, one can make use of the third moment of the Hillas ellipse along the length axis (skewness). The distance from the center of gravity of the Hillas shower ($\delta$, see central panel in Fig.~\ref{camera_fig}) can be estimated using lookup tables. However, it has to be noted that there is a strong ambiguity between the energy of the particle and its impact distance; e.g. a shower coming from a $\gamma$-ray with medium energy and moderate impact distance can be misinterpreted as originating from a $\gamma$-ray with large energy and impact distance. To sum up, monoscopic reconstruction allows to lower the energy threshold of the system at the expense of angular resolution, energy resolution, and $\gamma$-hadron separation.

\section{Expected Performance}
\label{performance}

The above-mentioned analysis chains were applied to simulated $\gamma$-rays, protons, and electrons with energies $> 10\,$GeV in order to test the analysis performance. The simulations were carried out for a zenith angle of $20^{\circ}$ and a wobble offset of $0.5^{\circ}$. An example of a simulated $\gamma$-ray with an energy of $125\,$GeV for this configuration is shown in the left panel of Fig.~\ref{camera_fig}.

In the following, we compare some of the performance parameters of H.E.S.S. II in monoscopic mode for the different analysis methods. The chains are still under development and not to be considered final; furthermore, all values were obtained from Monte Carlo simulations. For these two reasons, all results shown here have to be considered preliminary.

\subsection{Effective Area}

One of the key parameters that determines the performance of a reconstruction method is the effective area for detecting $\gamma$-rays. It is furthermore decisive for the energy threshold.
 \begin{figure}[t]
  \centering
  \includegraphics[width=0.45\textwidth]{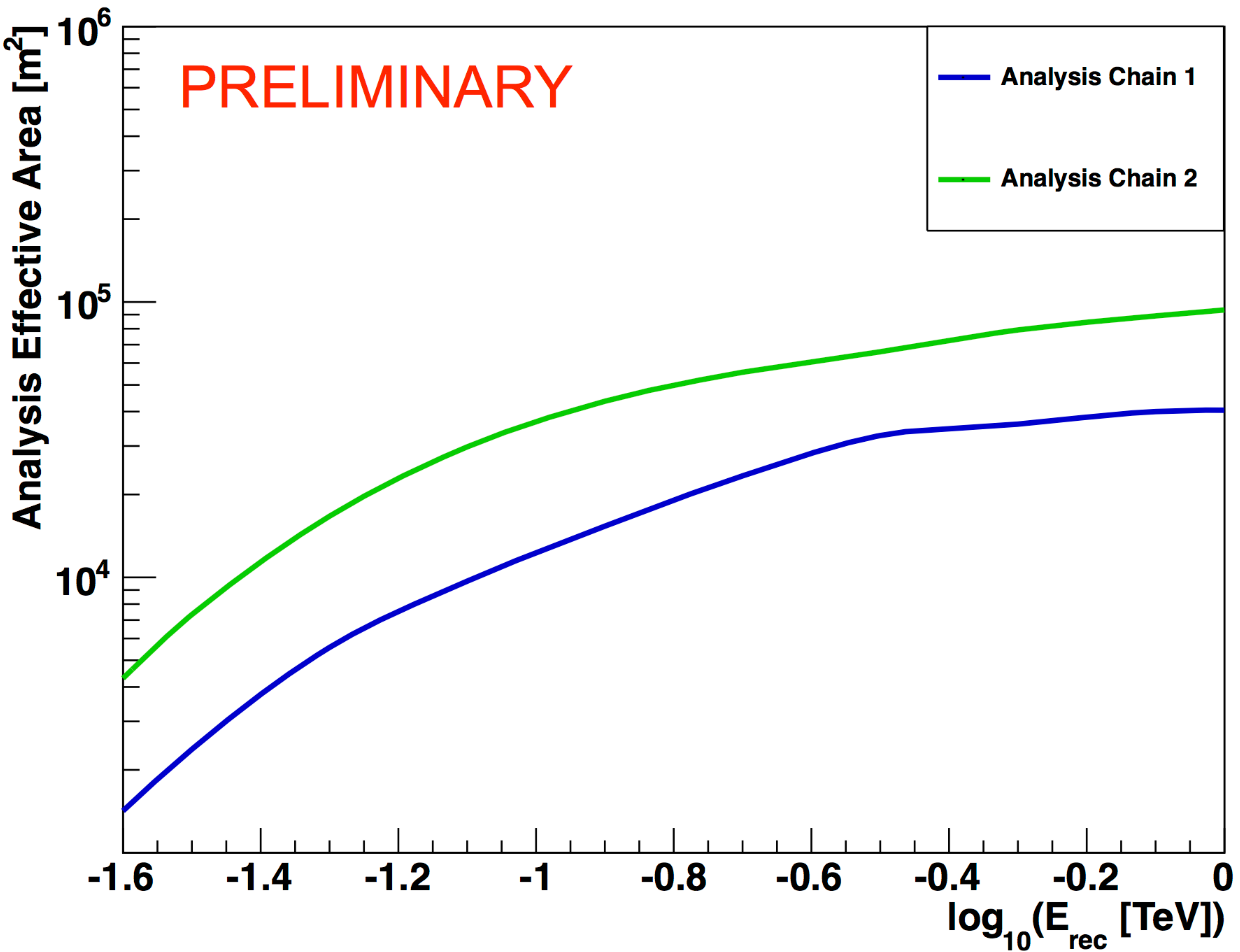}
  \caption{Effective Area for $\gamma$-rays as a function of the reconstructed energy.}
  \label{area_fig}
 \end{figure}
The analysis effective area for photons as a function of the reconstructed energy is shown in Fig.~\ref{area_fig} for both analysis chains. It reaches $10\%$ of its maximum value already at $\approx 30\,$GeV, reflecting a much lower energy threshold compared to H.E.S.S. I.  

\subsection{Integral Sensitivity}

In order to better characterize the performance of H.E.S.S. II in monoscopic mode, we calculated the sensitivity for detecting point-like $\gamma$-ray sources. For the calculation of the significance, we adopted the best-fit spectrum for the Crab Nebula from \cite{bib:CrabHess}. Assuming that the background is well-known, we estimated the significance as $N_{\mathrm{signal}} / \sqrt{N_{\mathrm{background}}}$ (following Eq.~(10b) from \cite{bib:LiMa}). The integral sensitivity $S_{\mathrm{int}}(E_{\mathrm{rec}})$ is defined as the flux of a source that reaches a significance of at least $5$ above a given reconstructed energy $E_{\mathrm{rec}}$ for an observation time of $50\,$h. In order to account for systematic uncertainties in the background determination, we required $N_{\mathrm{signal}} / N_{\mathrm{background}} > 0.01$. Additionally, the minimum number of signal counts $N_{\mathrm{signal}}$ was set to $10$. 

\begin{table}[h]
\begin{center}
\begin{tabular}{|l|c|}
\hline Analysis Chain & $S_{\mathrm{int}}$($100\,$GeV) \\ 
								  & [Crab]  \\ \hline
1 & $13.2\%$  \\ \hline
2 & $4.6\%$ \\ \hline
\end{tabular}
\caption{Integral sensitivities of the two analysis chains, assuming $1\%$ background systematics and requiring $N_{\mathrm{signal}} > 10$.}
\label{sens_table}
\end{center}
\end{table}
The integral sensitivity above $100\,$GeV for both chains is given in Table~\ref{sens_table} in units of the Crab Nebula flux. 

\subsection{Application to Real Data}
\label{crab}

We analyzed first data of the Crab Nebula taken with the new telescope using monoscopic events only. The data were taken at a zenith angle of $46^{\circ}$.
 \begin{figure}[t]
 \centering
 \includegraphics[width=0.45\textwidth]{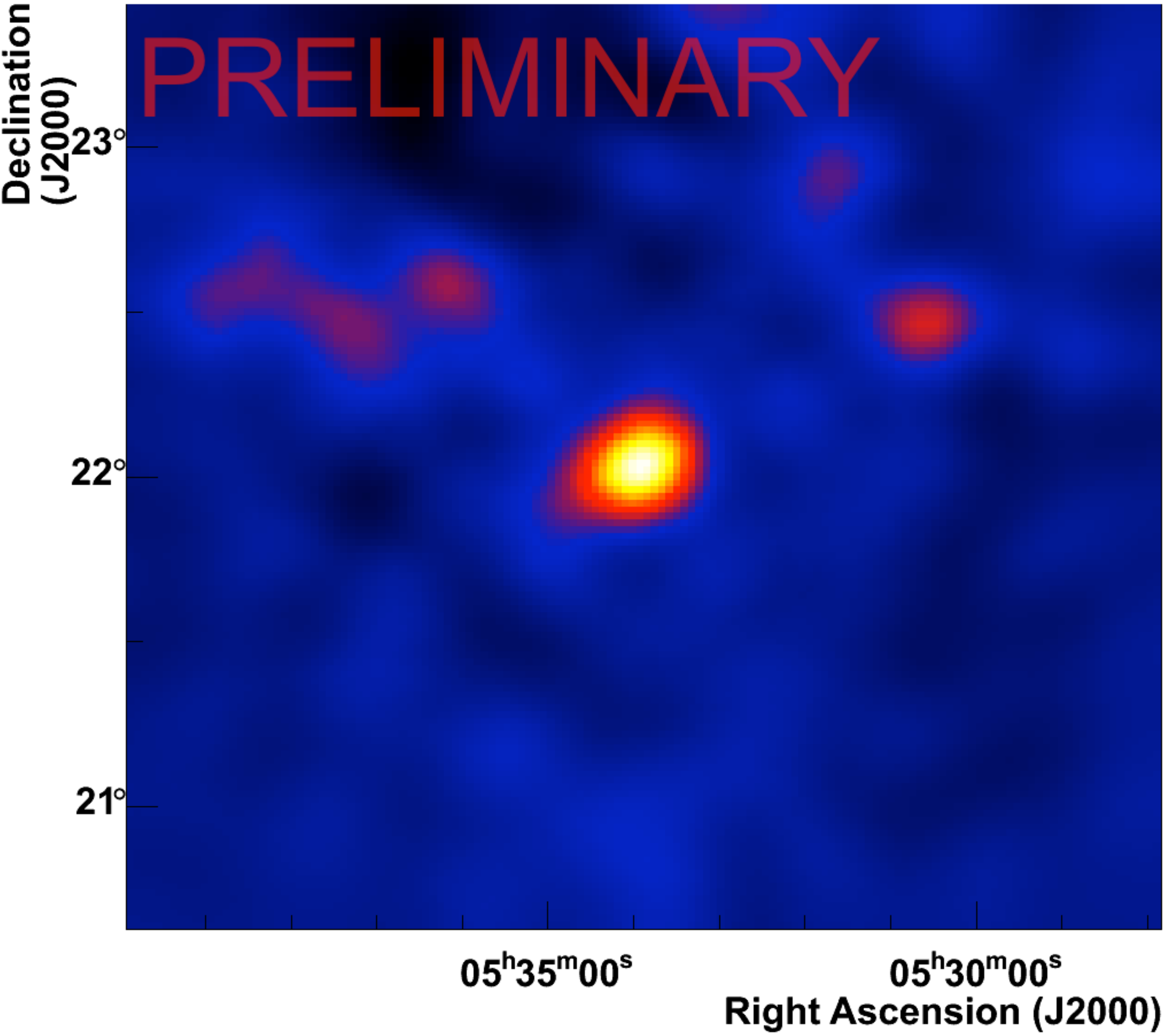}
 \caption{Excess map of the Crab Nebula as seen with the new telescope above $50\,$GeV using monoscopic events only, oversampled with a radius of $0.12^{\circ}$. The data have been analyzed using \textit{Chain 2}.}
  \label{crab_map}
 \end{figure}
 Fig.~\ref{crab_map} contains the excess map above $50\,$GeV obtained using the Analysis \textit{Chain 2}. The map was oversampled with a radius of $0.12^{\circ}$. The Crab Nebula is clearly visible, demonstrating the capabilities of the monoscopic reconstruction mode for the large H.E.S.S.~II telescope.

\section{Conclusions}
\label{concl}

We introduced two independent monoscopic analysis chains for H.E.S.S. II and compared their analysis effective areas and integral sensitivities. Since all performance results shown are based on Monte Carlo simulations, they have to be considered preliminary. 
We tested the monoscopic reconstruction technique on real H.E.S.S. II data of the Crab Nebula. The source appears as a distinct point-like source on the excess map at energies previously unaccessible to H.E.S.S., proving the potential of this reconstruction approach in order to close the energy gap to the Fermi satellite.

\footnotesize{{\bf Acknowledgements.}{ The support of the Namibian authorities and of the University of Namibia in facilitating the construction and operation of H.E.S.S. is gratefully acknowledged, as is the support by the German Ministry for Education and Research (BMBF), the Max Planck Society, the French Ministry for Research, the CNRS-IN2P3 and the Astroparticle Interdisciplinary Programme of the CNRS, the U.K. Particle Physics and Astronomy Research Council (PPARC), the IPNP of the Charles University, the South African Department of Science and Technology and National Research Foundation, and by the University of Namibia. We appreciate the excellent work of the technical support staff in Berlin, Durham, Hamburg, Heidelberg, Palaiseau, Paris, Saclay, and in Namibia in the construction and operation of the equipment.}}

\end{document}